\documentclass[reprint,superscriptaddress,floatfix,amsmath,amssymb,pra,aps,showpacs,twocolumn]{revtex4-1}
\usepackage{amsmath,amsfonts,amssymb,amsthm,graphics,graphicx,epsfig,bbm}
\usepackage[colorlinks=true,citecolor=blue,linkcolor=blue,urlcolor=blue]{hyperref}
\usepackage[usenames]{color}
\usepackage{graphicx}
\usepackage{subfigure}
\usepackage{amsmath}
\usepackage{epsfig}
\usepackage{dcolumn}
\usepackage{bm}
\usepackage{color}
\usepackage{times}
\usepackage{epstopdf}
\usepackage{amssymb}
\usepackage{amstext}
\usepackage{latexsym}
\usepackage{hyperref}
\usepackage{amsfonts}
\usepackage{psfrag}

\newcommand{\ket}[1]{\vert #1 \rangle}
\newcommand{\bra}[1]{\langle #1 \vert}

\newcommand{\braket}[2]{\langle #1 \vert #2 \rangle}

\newcommand{\abs}[1]{| #1 |}

\begin{document}

	\title{Measurement-based adaptation protocol with quantum reinforcement learning} 
	\date{\today}
	
	\author{F. Albarr\'an-Arriagada}
	\email[F. Albarr\'an-Arriagada]{\qquad francisco.albarran@usach.cl}
	\affiliation{Departamento de F\'isica, Universidad de Santiago de Chile (USACH), 
		Avenida Ecuador 3493, 9170124, Santiago, Chile}
	\affiliation{Center for the Development of Nanoscience and Nanotechnology 9170124, Estaci\'on Central, Santiago, Chile}
	
	\author{J. C. Retamal}
	\affiliation{Departamento de F\'isica, Universidad de Santiago de Chile (USACH), 
		Avenida Ecuador 3493, 9170124, Santiago, Chile}
	\affiliation{Center for the Development of Nanoscience and Nanotechnology 9170124, Estaci\'on Central, Santiago, Chile}
	
	\author{E. Solano}
	\affiliation{Department of Physical Chemistry, University of the Basque Country UPV/EHU, Apartado 644, 48080 Bilbao, Spain}
	\affiliation{IKERBASQUE, Basque Foundation for Science, Maria Diaz de Haro 3, 48013 Bilbao, Spain}
	\affiliation{Department of Physics, Shanghai University, 200444 Shanghai, China}
	\author{L. Lamata}
	\affiliation{Department of Physical Chemistry, University of the Basque Country UPV/EHU, Apartado 644, 48080 Bilbao, Spain}
	\begin{abstract}
		Machine learning employs dynamical algorithms that mimic the human capacity to learn, where the reinforcement learning ones are among the most similar to humans in this respect. On the other hand, adaptability is an essential aspect to perform any task efficiently in a changing environment, and it is fundamental for many purposes, such as natural selection. Here, we propose an algorithm based on successive measurements to adapt one quantum state to a reference unknown state, in the sense of achieving maximum overlap. The protocol naturally provides many identical copies of the reference state, such that in each measurement iteration more information about it is obtained. In our protocol, we consider a system composed of three parts, the ``environment'' system, which provides the reference state copies; the register, which is an auxiliary subsystem that interacts with the environment to acquire information from it; and the agent, which corresponds to the quantum state that is adapted by digital feedback with input corresponding to the outcome of the measurements on the register. With this proposal we can achieve an average fidelity between the environment and the agent of more than $90\% $ with less than $30$ iterations of the protocol. In addition, we extend the formalism to $ d $-dimensional states, reaching an average fidelity of around $80\% $ in less than $400$ iterations for $d=$ 11, for a variety of genuinely quantum and semiclassical states. This work paves the way for the development of quantum reinforcement learning protocols using quantum data and for the future deployment of semi-autonomous quantum systems.
	\end{abstract}
	
	\maketitle

\section{Introduction}
Machine learning (ML) is an area of artificial intelligence that focuses on the implementation of learning algorithms, and which has undergone great development in recent years~\cite{Russell1995Book, Michalski2013Book, Jordan2015}. ML can be classified into two broad groups, namely, learning by means of big data and learning through interactions. For the first group there are two classes, supervised learning, which uses previously classified data to train the learning program, inferring the function of relationship to classify new data. This is the case, e.g., of pattern recognition problems~\cite{Kawagoe1984, Shannon1995, Jain2007,Carrasquilla2017}. The other class is unsupervised learning, which does not require training data; instead, this learning paradigm uses the big data distribution to obtain an optimal method of classification using specific characteristics. An example is the clustering problem~\cite {Fahad2014,Baldi2014}. 

For the second group, learning from interactions, there is the class of reinforcement learning (RL)~\cite{Sutton1998Book, Littman2015}. RL is the learning paradigm most similar to the human learning process. Its general framework is as follows: we define two basic systems, an agent $A$ and an environment $E$, while often it is useful to define a register $R$ as an auxiliary system. The concept consists of $A$ inferring information by direct interaction with $E$, or indirectly, using the system $R$ as a mediator. With the information obtained, $A$ makes a decision to perform a certain task. If the result of this task is good, then the agent receives a reward, otherwise the agent receives a punishment. In addition, the RL algorithms can be divided into three basic parts, the policy, the reward function (RF), and the value function (VF). The policy can be subdivided into three stages: first, \textit{interaction with the environment}, in this stage, the way in which $A$ or $R$ interacts with $ E $ is specified; second, \textit{information extraction}, which indicates how $A$ obtains information from $E$; and finally, \textit{the action}, where $A$ makes the decision of what to do with the information of the previous step. RF refers to the criterion to assign the reward or punishment to $A$ in each iteration. And VF evaluates the utility of $A$ referred to the given task. An example of RL consists of artificial players for go or chess~\cite{Silver2017go, Silver2017Chess}.

 Another essential aspect of the RL protocols is the exploitation-exploration relation. Exploitation refers to the ability to make good decisions, while exploration is the possibility of making different decisions. For example, if we want to select a gym to do sports, the exploitation is given by the quality of the gym we test, while the exploration is the size of the search area in which we will choose a new gym to test. In the RL paradigm, a good exploitation-exploration relation can guarantee the convergence of the learning process, and its optimization depends on each algorithm.
\begin{figure}[b]
	\centering
	\includegraphics[width=.8\linewidth]{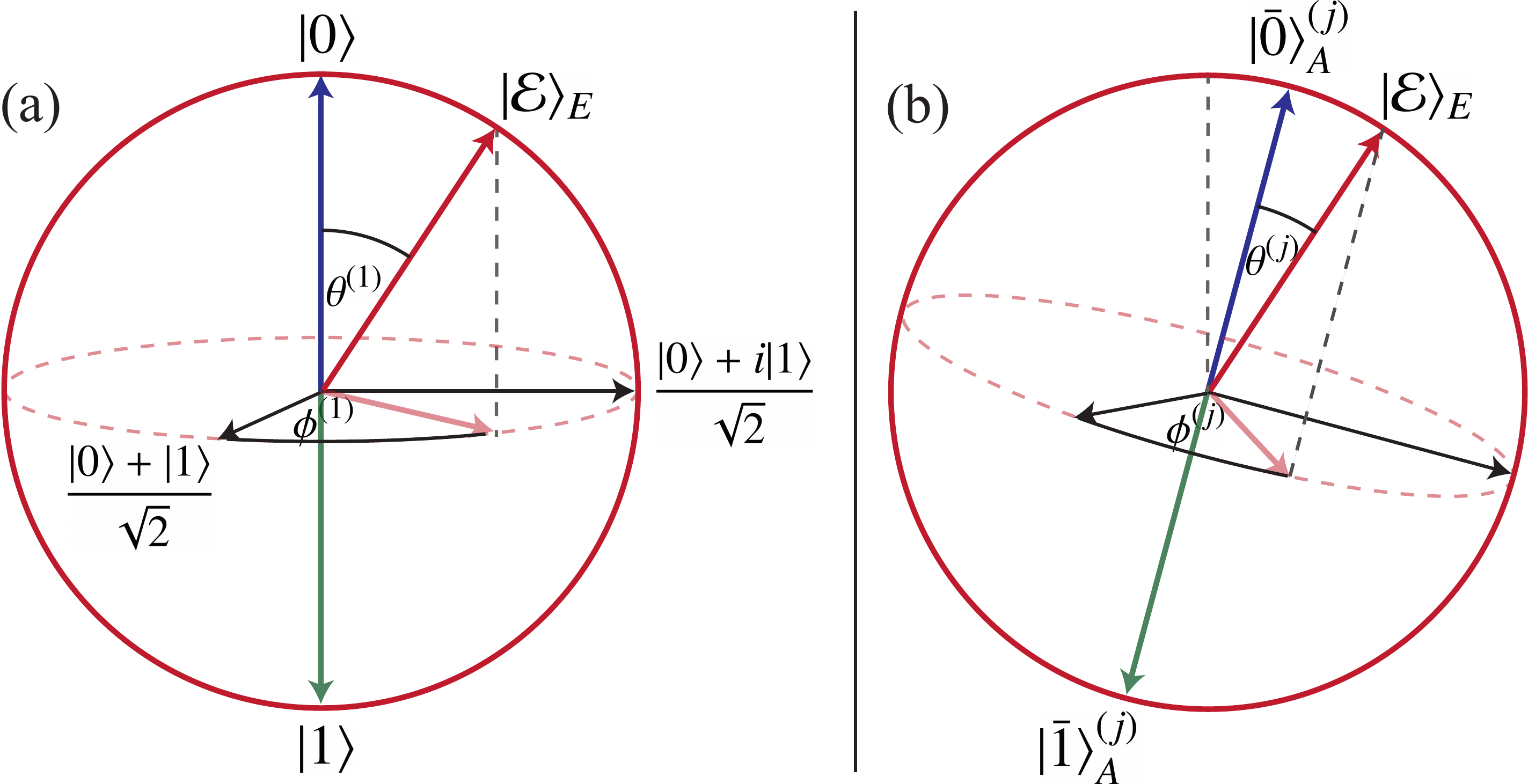}
	\caption{(a) Bloch representation of the environment state at the initial time, with $\ket{0}$ being the state of the agent. (b) Bloch representation of the environment in the $j$th iteration, where $\ket{\bar{0}_j}$ is the state of the agent, which was rotated in the previous iterations.}
	\label{BlochSphere}
\end{figure}

On the other hand, quantum mechanics is known to improve computational tasks~\cite{Nielsen2010Book}, so a natural question is the following: how are the learning algorithms modified in the quantum domain? To answer this question the quantum machine learning field (QML) has emerged. In recent years, QML has been a fruitful area~\cite{Schuld2015, Adcock2015,Dunjko2016, Dunjko2017, Biamonte2017,Biswas2017,PerdomoOrtiz2017,PerdomoOrtiz2017_2,Bukov2018}, in which quantum algorithms have been developed~\cite{Sasaki2002, Lloyd2013,Benedetti2017,Benedetti2017_2} that show a possible speed-up in certain situations in relation with their classical counterparts~\cite{Aimeur2013, Paparo2014}. However, these novel works focus mainly on learning from classical data encoded in quantum systems, processed with a quantum algorithm and decoded to be read by a classical machine. In this context, the speed-up of quantum machine learning is often balanced with the necessary resources to encode and decode the information, which leads to an unclear quantum supremacy. Nevertheless, recent works analyze the QML paradigm in a purely quantum way~\cite{Lamata2017, CardenasLopez2017, AlvarezRodriguez2017}, in which quantum systems learn quantum data.

In this context, one of the most fundamental tasks is the learning of a quantum state that describes a quantum system. In this case, there are many protocols related to quantum state estimation \cite{Adamson2010,SosaMartinez2017,Lumino2017,Rocchetto2017,Torlai2017,Sun2018}, but in these cases, the quantum data are learned by a classical system. Therefore, to encode this information in another quantum system we need many resources. Another approach is approximate quantum cloning \cite{Scarani2005}, where we need some information about the basis of the state to clone and specific interactions, which hinders autonomous learning of any quantum state by a quantum system and limits the architectures to implement it. Therefore, a protocol in which a quantum system can learn an arbitrary state of another quantum system without additional information, in order to obtain an autonomous quantum device, is an open problem. Furthermore, this approach is valuable when we want to use only available resources in any quantum technology platform, such as measurements and feedback loop processes.
	
In this article, we present a quantum machine learning algorithm based on a reinforcement learning approach, to adapt the quantum state of a system (agent), to the unknown state of another quantum system (environment), assisted by the measurements on a third system (register). This algorithm uses multiple identical copies of the Environment which are entangled to the Register. We propose to use coherent feedback loops, conditioned to the measurements in order to perform the adaptation process without human intervention. Therefore, this approach is different from projective simulation \cite{Briegel2012,Tiersch2015,Mautner2015,Melnikov2017}, which makes use of a set of projective measurements with different probabilities at the same time.

In our numerical calculations we obtain average fidelities of more than $90\%$ for qubit states after less than $40$ measurements, while for qudits the protocol achieves average fidelities of $80\%$ using $400$ iterations with $d=11$ dimensions, for either genuinely quantum states or semiclassical states. This proposal can be useful for the implementation of semiautonomous quantum devices.
	
\section{The Quantum Adaptation Algorithm}
Our framework is as follows. We assume a known quantum system called the agent ($A$) and many copies of an unknown quantum state provided by a system called the environment ($E$). We also consider an auxiliary system called the register ($R$) which interacts with $E$. Then, we obtain information about $E$ by measuring $R$, and we employ the result as an input to the RF function. Finally, we perform a partially random unitary transformation on $A$, which depends on the output of the RF. The idea is to improve the fidelity between $A$ and $E$, without projecting the state of $A$ with measurements. 

This algorithm differs from quantum cloning because we use a pseudorandom paradigm to obtain a balance in the exploration-exploitation relation without additional information about the environment. This protocol also differs from quantum state estimation in the fact that we propose a semiautonomous quantum agent; that is, the aim is that in the future a quantum agent will learn the state of the environment without any human intervention. Other authors have considered the inverse problem, an unknown state evolved to a known state assisted by measurements \cite{Roa2006}, which deviates from the machine learning paradigm. Therefore, an optimal measurement is not performed in each step, but after a certain number of autonomous iterations, the agent converges to a large fidelity with the unknown state.

In the rest of the article we use the following notation: the subscripts $A$, $R$, and $E$ refer to each subsystem, and the superscripts indicate the iteration. For example, $\mathcal{O}_{\alpha}^{(k)}$ refers to the operator $\mathcal{O}$ that acts on the subsystem $\alpha$ during the $k$th iteration. Moreover, the lack of any of these indices indicates that we are referring to a general object in the iterations and/or in the subsystems.

We start with the case where each subsystem is described by a qubit state. We assume that $A(R)$ is described by $\ket{0}_{A(R)}$, and $E$ is described by an arbitrary state expressed in the Bloch sphere as $\ket{\mathcal{E}}_E=\cos(\theta^{(1)}/2)\ket{0}_E+e^{-i\phi^{(1)}}\sin(\theta^{(1)}/2)\ket{1}_E$ [see Fig.~\ref{BlochSphere} (a)]. The initial state reads
	\begin{equation}
	\ket{\psi^{(1)}}=\ket{0}_A\ket{0}_R[\cos(\theta^{(1)}/2)\ket{0}_E+e^{i\phi^{(1)}}\sin(\theta^{(1)}/2)\ket{1}_E].
	\label{qbPsi0}
	\end{equation}
First of all, we introduce the general elements of our reinforcement learning protocol, such as the policy, the RF, and the VF. For the policy, we perform a controlled-NOT (CNOT) gate ($U^{NOT}_{E,R}$) with $E$ as the control and $R$ as the target (i.e., the interaction with the environment), in order to copy information of $E$ into $R$, obtaining 
	\begin{eqnarray}
	\ket{\Psi_1}=U^{NOT}_{E,R}\ket{\psi^{(1)}}&&=\ket{0}_A\Big[\cos(\theta^{(1)}/2)\ket{0}_R\ket{0}_E\nonumber\\
	&&+e^{i\phi^{(1)}}\sin(\theta^{(1)}/2)\ket{1}_R\ket{1}_E\Big].
	\end{eqnarray}
We then measure the register qubit in the basis $ \{\ket{0},\ket{1}\}$, with the probability $p_0^{(1)}=\cos^2(\theta_1 /2)$ or $p_1^{(1)}=\sin^2 (\theta_1/2)$, to obtain the state $ \ket{0}$ or $ \ket{1}$, respectively (i.e., information extraction). If the result is $ \ket{0}$, it means that we have collapsed $ E $ into $ A $ and do nothing, but if the result is $\ket{1}$, it means that we have measured the component of $E$ orthogonal to $A$, and thus we accordingly modify the agent. As we do not have additional information about the environment, we perform a partially-random unitary operator on $ A $ given by $U_A^{(1)}(\alpha^{(1)},\beta^{(1)})=e^{-iS^{z(1)}_A\alpha^{(1)}} e^{-iS^{x(1)}_A\beta^{(1)}}$ (action), where $\alpha^{(1)}$ and $\beta^{(1)}$ are random angles of the form $\alpha(\beta)^{(1)}=\xi_{\alpha(\beta)}\Delta^{(1)}$, where $\xi_{\alpha(\beta)}\in[-1/2,1/2]$ is a random number, $\Delta^{(1)}$ is the range of random angles, $\alpha(\beta)^{(1)} \in [-\Delta^{(1)}/2,\Delta^{(1)}/2]$, and $S^{k (1)}_A=S^{k}$ is the $k$th spin component. Now, we initialize the register qubit state and employ a new copy of $E$, obtaining the next initial state for the second iteration:
	\begin{equation}
	\ket{\psi^{(2)}}=\mathcal{U}_A^{(1)}\ket{0}_A\ket{0}_R\ket{\mathcal{E}}_E=\ket{\bar{0}}^{(2)}_A\ket{0}_R\ket{{\mathcal{E}}}_E,
	\end{equation}
with 
	\begin{equation}
	\mathcal{U}_A^{(1)}=\Big[m^{(1)}U_A^{(1)}(\alpha^{(1)},\beta^{(1)})+(1-m^{(1)})\mathbb{I}_A\Big].
	\label{change}
	\end{equation}
Here $m^{(1)}=\{0,1\}$ is the outcome of the measurement, $\mathbb{I}$ is the identity operator, and we define the new agent state as $\ket{\bar{0}}_1^{(2)}=\mathcal{U}_1^{(1)}\ket{0}_1$.
	
Now, we define the RF to modify the exploration range of the $k$th iteration $\Delta^{(k)}$ as
	\begin{equation}
	\Delta^{(k)}=\bigg[(1-m^{(k-1)})\mathcal{R}+m^{(k-1)}\mathcal{P}\bigg]\Delta^{(k-1)},
	\label{RewarFunction}
	\end{equation} 
where $m^{(k-1)}$ is the outcome of the $(k-1)$th iteration, while $\mathcal{R}$ and $\mathcal{P}$ are the reward and punishment ratios, respectively. Equation (\ref{RewarFunction}) means that the value of $\Delta$ is modified by $\mathcal{R}\Delta$ for the next iteration when $m=0$ and by $\mathcal{P}\Delta$ when the outcome is $m=1$. In our protocol, we choose for simplicity $\mathcal{R}=\epsilon<1$ and $\mathcal{P}=1/\epsilon>1$, such that, every time that the state $\ket{0}$ is measured, the value of $\Delta$ is reduced, and it is increased in the other case. Also, the fact that $\mathcal{R}\cdot\mathcal{P}=1$ means that the punishment and the reward have the same strength, or in other words, if the protocol yields the same number of outcomes $0$ and $1$, the exploration range does not change. Finally, the VF is defined as the value of $\Delta^{(n)}$ after all iterations. Therefore, $\Delta^{(n)}\rightarrow0$ if the protocol improves the fidelity between $A$ and $E$.

\begin{figure}[t]
	\centering
	\includegraphics[width=.7\linewidth]{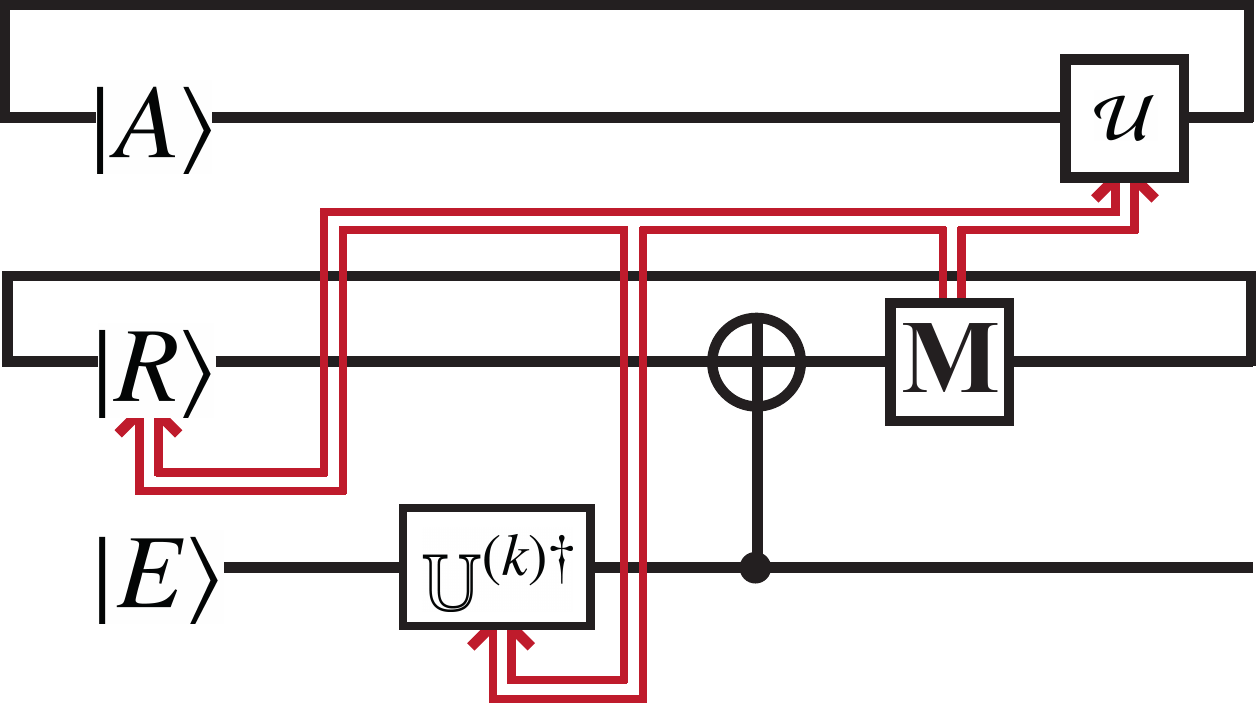}
	\caption{Quantum circuit diagram for the measurement-based adaptation protocol. The box labelled with M indicates the projective measurement process, and the red lines denote feedback loops.}
	\label{Circuit}
\end{figure}

To illustrate how the protocol proceeds, let us consider the $k$th iteration. The initial state is given by

	\begin{equation}
	\ket{\psi}^{(k)}=\ket{\bar{0}}_A^{(k)}\ket{0}_R\ket{\mathcal{E}}_E,
	\end{equation}
where $\ket{\bar{0}}_A^{(k)}=\mathbb{U}_A^{(k)}\ket{0}_A$ and $\mathbb{U}_A^{(k)}=\mathcal{U}_A^{(k-1)}\mathbb{U}_A^{(k-1)}$, where $\mathbb{U}_A^{(1)}=\mathbb{I}_A$ and $\mathcal{U}_A^{(j)}$ is given by Eq.~(\ref{change}). Also $U_A^{(j)}=e^{-iS_A^{z(j)}\alpha^{(j)}}e^{-iS_A^{x(j)}\beta^{(j)}}$, where we define 
\begin{eqnarray}
&&S_A^{z(j)}=\frac{1}{2}(\ket{\bar{0}}_A^{(j)}\bra{\bar{0}}-\ket{\bar{1}}_A^{(j)}\bra{\bar{1}})=\mathcal{U}_A^{(j-1)\dagger}S_A^{z(j-1)}\mathcal{U}_A^{(j-1)},\nonumber\\ &&S_A^{x(j)}=\frac{1}{2}(\ket{\bar{0}}_A^{(j)}\bra{\bar{1}}+\ket{\bar{1}}_A^{(j)}\bra{\bar{0}})=\mathcal{U}_A^{(j-1)\dagger}S_A^{x(j-1)}\mathcal{U}_A^{(j-1)},
\label{Sj}\nonumber\\
\end{eqnarray}
with ${_A^{(j)}}\braket{\bar{0}}{\bar{1}}_A^{(j)}=0$. We can write the state of $E$ in the Bloch representation using $\ket{\bar{0}_j}$ as a reference axis [see Fig.~\ref{BlochSphere} (b)] and apply the operator $\mathbb{U}^{(k)\dagger}_E$, obtaining the following for $E$,
\begin{eqnarray}
&&\mathbb{U}^{(k)\dagger}_E\ket{\mathcal{E}}_E\nonumber\\
&&=\mathbb{U}^{(k)\dagger}_E\bigg[\cos(\theta^{(k)}/2)\ket{\bar{0}}^{(k)}_E+e^{i\phi^{(k)}}\sin(\theta^{(k)}/2)\ket{\bar{1}}^{(k)}_E\bigg]\nonumber\\
&&=\cos(\theta^{(k)}/2)\ket{0}_E+e^{i\phi^{(k)}}\sin(\theta^{(k)}/2)\ket{1}_E=\ket{\bar{\mathcal{E}}}_E^{(k)}.\quad
\end{eqnarray}
We can write the states $\ket{\bar{0}^{(k)}}$ and $\ket{\bar{1}^{(k)}}$ in terms of the initial logical states $\ket{0}$ and $\ket{1}$ and the unknown angles $\theta^{(k)}$, $\theta^{(1)}$, $\phi^{(k)}$, and $\phi^{(1)}$ as follows:
\begin{eqnarray}
\ket{\bar{0}}^{(k)}=&&\cos\Bigg(\frac{\theta^{(1)}-\theta^{(k)}}{2}\Bigg)\ket{0}+e^{i\phi^{(1)}}\sin\Bigg(\frac{\theta^{(1)}-\theta^{(k)}}{2}\Bigg)\ket{1},\nonumber\\
\ket{\bar{1}}^{(k)}=&&-e^{-i\phi^{(k)}}\sin\Bigg(\frac{\theta^{(1)}-\theta^{(k)}}{2}\Bigg)\ket{0}\nonumber\\
&&+e^{i(\phi^{(1)}-\phi^{(k)})}\cos\Bigg(\frac{\theta^{(1)}-\theta^{(k)}}{2}\Bigg)\ket{1}.
\end{eqnarray}
Therefore, the operator $\mathbb{U}^{(k)\dagger}$ performs the necessary rotation to transform $\ket{\bar{0}^{(k)}}\rightarrow\ket0$ and $\ket{\bar{1}^{(k)}}\rightarrow\ket1$. Then, we perform the gate $U^{NOT}_{E,R}$,
\begin{eqnarray}
\ket{\Phi^{(k)}}=&&U^{NOT}_{E,R}\ket{\bar{0}}^{(k)}_A\ket{0}_R\ket{\bar{\mathcal{E}}}_E\nonumber\\
=&&\ket{\bar{0}}_A^{(k)}\bigg[\cos(\theta^{(k)}/2)\ket{0}_R\ket{0}_E\nonumber\\
&&+e^{i\phi^{(k)}}\sin(\theta^{(k)}/2)\ket{1}_R\ket{1}_E\bigg],
\end{eqnarray}
and we measure $R$, with probabilities $p_0^{(k)}=\cos^2(\theta^{(k)}/2)$ and $p_1^{(k)}=\sin^2(\theta^{(k)}/2)$ for the outcomes $m^{(k)}=0$ and $m^{(k)}=1$, respectively. Finally, we apply the RF given by Eq.~(\ref{RewarFunction}). We point out that, probabilistically, when $p_0^{(k)}\rightarrow1$, $\Delta\rightarrow0$, and when $p_1^ {(k)}\rightarrow1$, $\Delta\rightarrow4\pi $. In terms of the exploitation-exploration relation, this means that when the exploitation decreases (we measure $\ket{1}$ often), we increase the exploration (we increase the value of $\Delta$) to increase the probability of making a beneficial change, and when the exploitation improves (we measure $\ket{0}$ many times), we reduce the exploration to allow only small changes in the following iterations. The diagram of this protocol is shown in Fig.~\ref{Circuit}.
	  
Figure~\ref{Graf2}(a) shows the numerical calculation of mean fidelity between $A$ and $E$ for the single-qubit case. For this computation we use $2000$ random initial states with $\epsilon=0.1$ ({blue-crosses line), $\epsilon=0.3$ (red-circles line), $\epsilon=0.5$ (yellow-pluses line), $\epsilon=0.7$ (purple-angle line), and  $\epsilon=0.9$ (green solid line). We can see that the protocol can reach fidelities over $90\%$ in less than $30$ iterations. Figure ~\ref{Graf2}(b) depicts the evolution of the exploration parameter $\Delta$ for each iteration for the same values of the constant $\epsilon$. We can see from Fig.~\ref {Graf2} that, when the parameter $\epsilon$ is small, the fidelity between $A$ and $E$ increases quickly (the learning speed increases), requiring less iterations to reach high fidelities; however, the maximum value of the average fidelity (maximum learning) is smaller than when $\epsilon$ increases. This means that small changes in the scan parameter $\Delta$ (large $\epsilon$) result in a higher but slower learning.

\section{Multilevel protocol}
In this section, we extend the previous protocol to the case where $A$, $R$, and $E$ are described by one $d$-dimensional qudit state. One of the ingredients in the qubit case is the CNOT gate. Here, we use the extension of the CNOT gate to multilevel states, also known as the XOR gate~\cite{Alber2000} ($U^{XOR}_{a,b}$). The action of this gate is given by
\begin{equation}
U^{XOR}_{a,b}\ket{j}_a\ket{k}_b=\ket{j}_a\ket{j\ominus k}_b,
\label{XOR}
\end{equation}
where the index $a$ ($b$) refers to control(target) state, and $\ominus$ denotes the difference modulo $d$, with $d$ being the dimension of each subsystem. The CNOT gate has two important properties, namely, (i) $U^{NOT}_{a,b}$ is Hermitian, and (ii) $U^{NOT}_{a,b}\ket{j}_a\ket{k}_b=\ket{j}_a\ket{0}_b$ if and only if $j=k$. These two properties are maintained in the XOR gate defined in Eq.~(\ref{XOR}). The policy and the VF are essentially the same as in the previous case, but now we consider the multiple outcomes ($m^{(j)}\in\{0,1,\dots,d-1\}$) that result from measuring $R$. First, we introduce $\ket{\bar{1}_j}_A=\ket{m^{(j)}}_A$ for the definition of $S^{z(j)}_A$ and $S^{x(j)}_A$ in Eq.~(\ref{Sj}). As in the previous case, we assume the initial state of $A$ to be $\ket{0}_A$, while $R$ is initialized in $\ket{0}_R$. Moreover, the state of $E$ is arbitrary and expressed as $\ket{\mathcal{E}}_E=\sum_{j=0}^{d-1}c_j\ket{j}_E$, where $\sum_{j=0}^{d-1}\abs{c_j}^2=1$, and $d$ is the dimension of $E$. We can rewrite $E$ in a more convenient way as
\begin{equation}
\ket{\mathcal{E}}_E=\cos(\theta^{(1)}/2)\ket{\bar{0}^{(1)}}_E+e^{i\phi^{(1)}}\sin(\theta^{(1)}/2)\ket{\bar{0}_{\perp}^{(1)}}_E,
\end{equation}
where $\ket{\bar{0}^{(1)}}_E=\ket{0}_E$, $\ket{\bar{0}_{\perp}^{(1)}}_E=(1/\mathcal{N})\sum_{j=1}^{d-1}c_j\ket{j}_E$ is the orthogonal component to $\ket{0}_E$, and $\mathcal{N}^2=\sum_{j=1}^{d-1}\abs{c_j}^2$. Subsequently, we perform the XOR gate $U^{XOR}_{E,R}$, obtaining
\begin{figure}[t]
	\centering
	\includegraphics[width=1\linewidth]{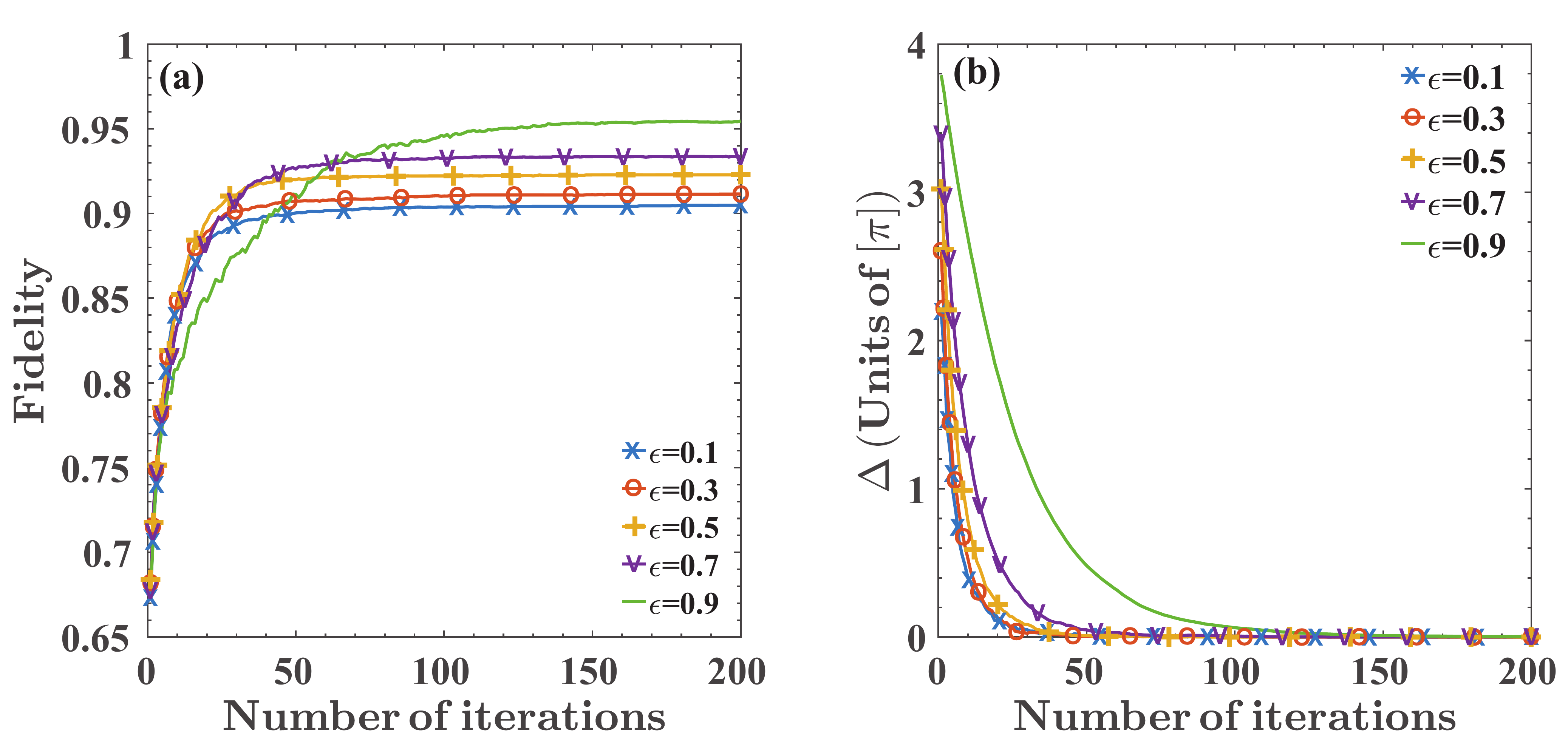}
	\caption{Performance of the measurement-based adaptation protocol for the qubit case. Panel (a) shows the mean fidelity for $2000$ initial random states, and panel (b) the value of $\Delta^{(j)}$ in each iteration.}
	\label{Graf2}
\end{figure}
\begin{eqnarray}
&&\ket{\Phi_0}=U^{XOR}_{E,R}\ket{0}_A\ket{0}_R\ket{\mathcal{E}}_E\nonumber\\
&&=\ket{0}_A\Big[\cos(\theta^{(1)}/2)\ket{0}_R\ket{0}_E+e^{i\phi^{(1)}}\sin(\theta^{(1)}/2)\ket{\varkappa}_{R,E}\Big],\nonumber\\
\end{eqnarray}
with $\ket{\varkappa}_{R,E}=\sum_{j=1}^{d-1}(1/\mathcal{N})c_j\ket{d-j}_R\ket{j}_E$. As in the previous case, we measure $R$, but now we have multiple outcomes and we separate them into two groups: first, the outcome $\ket{0}$ with the probability $p_0^{(1)}=\cos^2(\theta_1/2)$, and second, outcomes $\ket{j}$ with $j\ne0$, and the probability to obtain any of them of $p_{\perp}^{(1)}=\sin^2(\theta^{(1)}/2)$. As in the previous case, this means that we measure either in the state of $A$ or in the orthogonal subspace. With this information, we perform a partially random unitary operation on the agent $\mathcal{U}^{(1)}_A=e^{-iS^{z(1)}_A\alpha^{(1)}}e^{-iS^{x(1)}_A\beta^{(1)}}$, using the definition (\ref{Sj}) with $\ket{\bar{1}_A}=\ket{m^{(1)}}_A$, where $m^{(1)}=j$ is the outcome of the measurement. If $m^{(1)}=0$, then $\mathcal{U}^{(1)}_A=\mathbb{I}_A$. The random angles $\alpha^{(1)}$ and $\beta^{(1)}$ are defined as in the qubit case. Now, the RF changes slightly and is given by
\begin{equation}
\Delta^{(j)}=\bigg[\delta_{m^{(j-1)},0}\mathcal{R}+(1-\delta_{m^{(j-1)},0})\mathcal{P}\bigg]\Delta^{(j-1)},
\label{RFmulti}
\end{equation}
where $\delta_ {j, k}$ is the delta function. Equation (\ref {RFmulti}) means that if we measure $\ket{0}$ in $R$, the value of $\Delta$ decreases for the next iteration, and if we measure $\ket{j}$ with $j\ne0 $, $ \Delta$ increases. Remember that $\mathcal{R}=\epsilon<1$ and $\mathcal{P}=1/\epsilon>1$. As in the qubit case, the RF is binary, since all the results $\ket{j}$ with $j\ne0 $ are equally nonbeneficial, so we give the same punishment to the agent. For this reason we use the same policy as in the qubit protocol for the case of multiple levels. As in the case of a single-qubit state, the parameter $\epsilon$ plays a fundamental role in the learning process by handling the speed of learning and the maximum learning. This will be understood in what follows.

In what follows we study the performance of the protocol for $d$-dimensional states to describe $E$. As an example, we consider a random superposition of the form
\begin{figure}[t]
	\centering
	\includegraphics[width=1\linewidth]{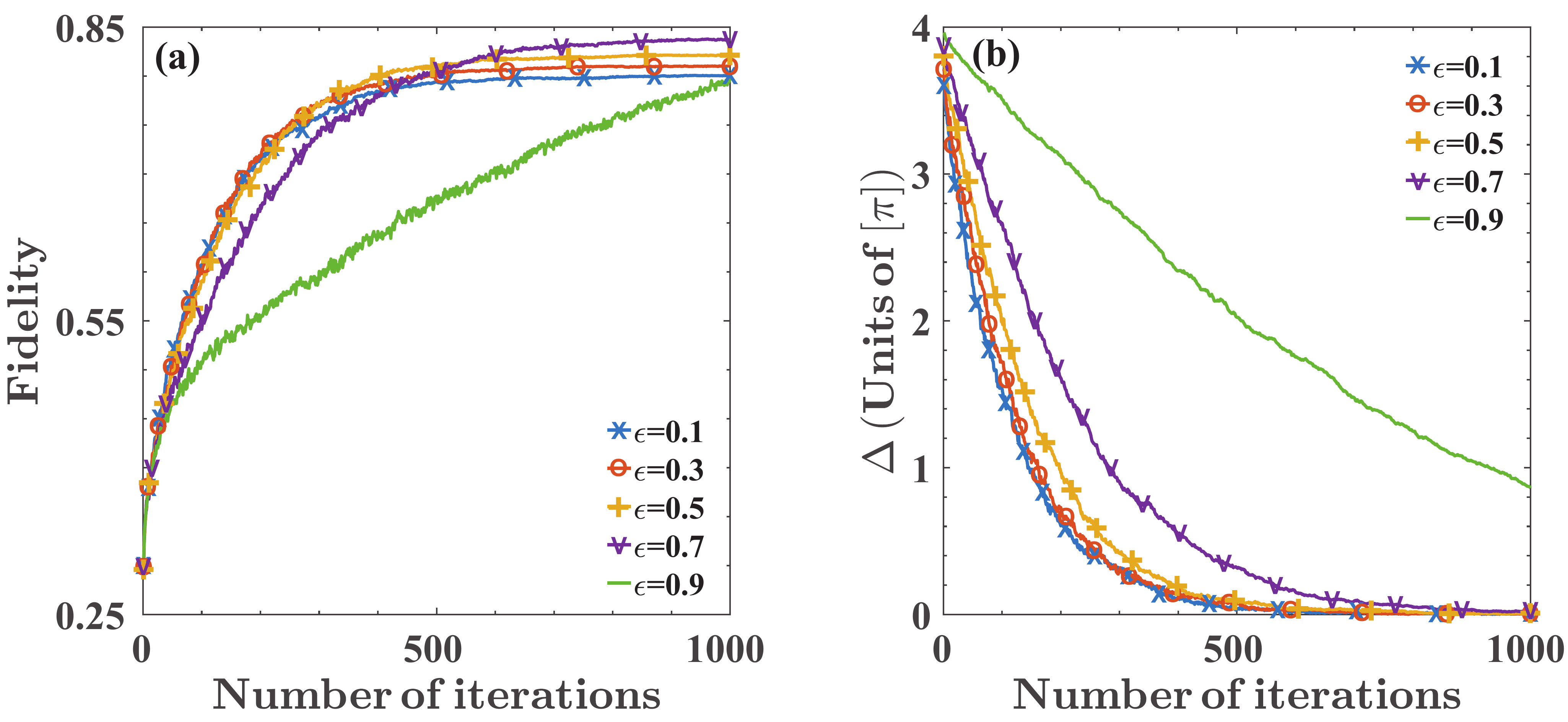}
	\caption{Performance of the measurement-based adaptation protocol for a total random state given by Eq.~(\ref{rand10}) with $d=11$. Panel (a) shows the mean fidelity for $2000$ initial random states, and panel (b) the value of $\Delta_j$ in each iteration.}
	\label{Graf10}
\end{figure}
\begin{figure}[b]
	\centering
	\includegraphics[width=1\linewidth]{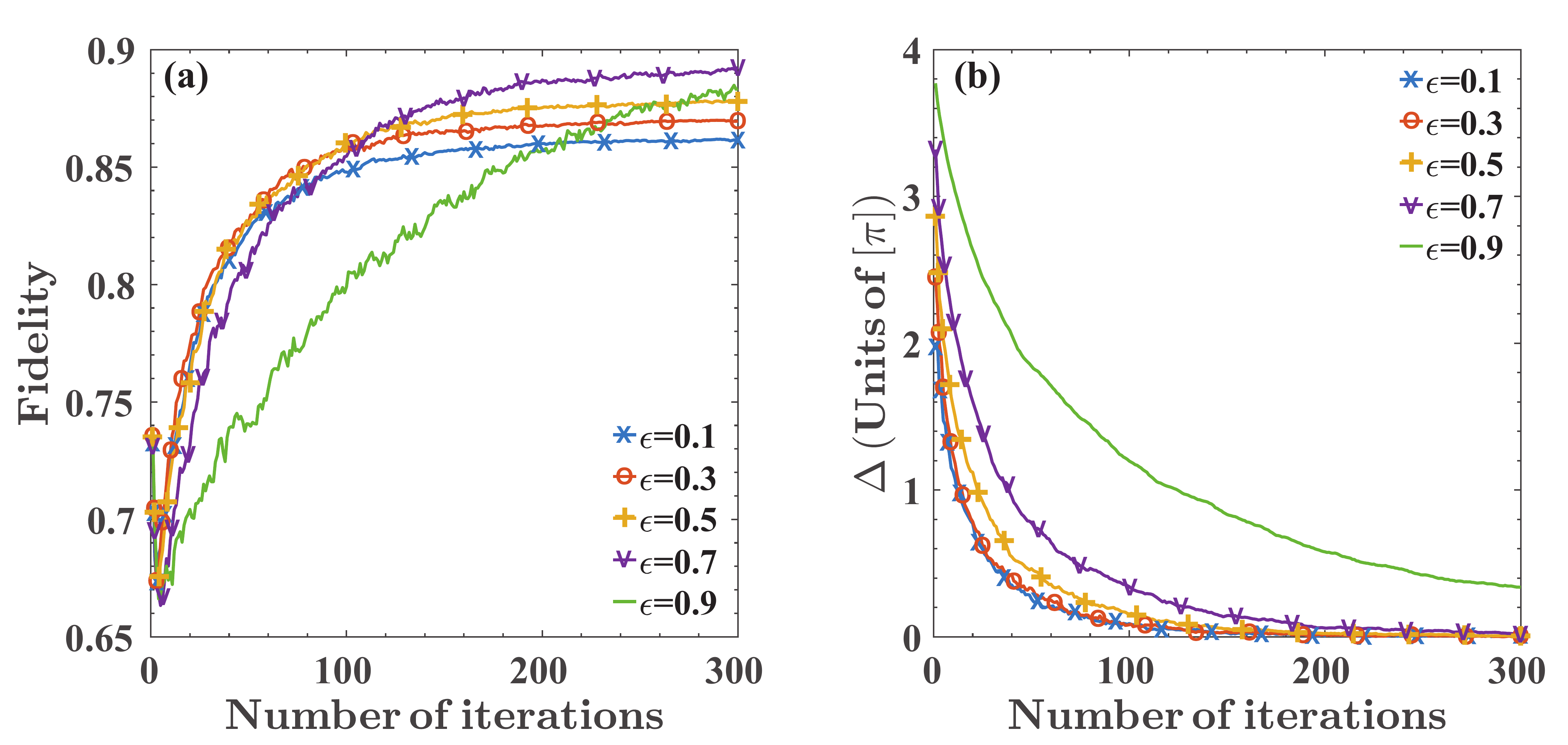}
	\caption{Performance of the measurement-based adaptation protocol for a coherent state of the form in Eq.~(\ref{Coherent}). Panel (a) shows the mean fidelity for $2000$ random pairs $\{a,b\}\in[0,1)$, where $\alpha=a+ib$; and panel (b) the value of $\Delta_j$ in each iteration.}
	\label{Coherent10}
\end{figure}
\begin{eqnarray}
\ket{\mathcal{E}}_E=\frac{1}{N}\sum_{k=0}^{d-1}c_k\ket{k}_E,\qquad c_k=a+ib,
\label{rand10}
\end{eqnarray}
where $a,b\in[0,1)$ are random numbers and $N$ is a normalization factor. Figure \ref{Graf10} shows the numerical calculations for the particular case of $d=11$, where panel (a) gives the average fidelity for $2000$ initial states given by Eq.~(\ref{rand10}), and panel (b) gives the evolution of $\Delta$ in each iteration. It also shows how this exploration parameter is reduced when the fidelity between $E$ and $A$ grows (increasing the exploitation). We can see from Fig. \ref{Graf10}(a) that the protocol can reach mean fidelities of $80\%$ within about $400$ iterations or, equivalently, the protocol increases the mean fidelity between $A$ and $E$ in about $0.5$ using $400$ iterations.

Although with the previous case we cover a large variety of $d$-dimensional states, let us implement the protocol for some standard quantum states in $d$ dimensions, which could be suitable for experimental realization. To this purpose, we consider in a first instance a coherent state defined by
\begin{equation}
\ket{\alpha}=e^{-\abs{\alpha}^2/2}\sum_{n=0}^{\infty}\frac{\alpha^n}{\sqrt{n!}}\ket{n}.
\label{Coherent}
\end{equation}
For this case we use $\alpha=a+ib$, with $a$ and $b$ being positive real random numbers smaller than $1$. As $\abs{\alpha}^2\leq2$, we can truncate the sum (\ref{Coherent}) to $n=10$, since the probabilities to obtain $\ket{n}$ with $n>10$ are bounded by $\abs{e^{-\abs{\alpha}^2/2}\alpha^{10}/\sqrt{10!}}^2\leq e^{-1}\sqrt{2}^{10}/\sqrt{10!}\approxeq0.0062$. Figure \ref{Coherent10}(a) shows the fidelity between $A$ and $E$ for each iteration, reaching values of $85\%$ in less than $100$ iterations. Figure \ref{Coherent10}(b) depicts the value of $\Delta$ in this process. We can also observe that the exploration is reduced when $A$ approaches $E$ (increasing the exploitation), as in the previous case.

\begin{figure}[t]
	\centering
	\includegraphics[width=1\linewidth]{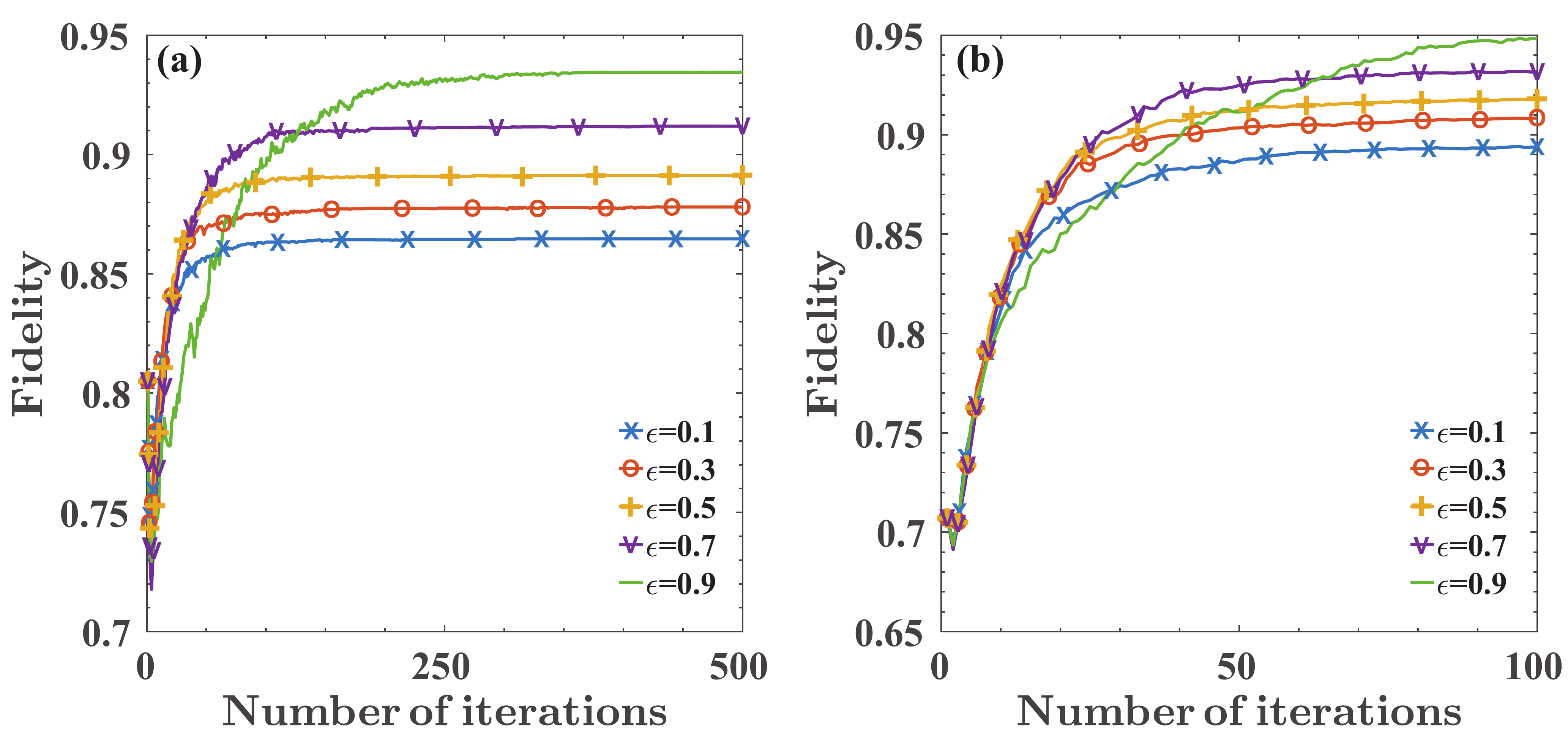}
	\caption{Performance of the measurement-based adaptation protocol for genuinely quantum states. Panel (a) shows the mean fidelity for $2000$ cat states of the form in Eq.~(\ref{CatState}), and panel (b) the mean fidelity for $2000$ repetitions of the protocol using the superposition $(\ket{0}_E+\ket{10}_E)/\sqrt{2}$ for the environment.}
	\label{CatON}
\end{figure}
In a second instance, we consider a superposition of two coherent states given by
\begin{equation}
\ket{\mathcal{E}}_E=\sqrt{\frac{1}{\mathcal{N}_{\alpha}}}\bigg(\ket{\alpha}+\ket{-\alpha}\bigg),
\label{CatState}
\end{equation}
which is known as a cat state. Here $\ket{\alpha}$ is given by Eq. (\ref{Coherent}) and $\mathcal{N}_{\alpha}$ is a normalization factor. And finally, we consider the superposition
\begin{equation}
\ket{\mathcal{E}}_E=\sqrt{\frac{1}{2}}\bigg(\ket{0}+\ket{n}\bigg).
\label{ONState}
\end{equation}
This state allows us to compare the performance of the protocol with the case of one qubit. We consider in particular the case of $n=10$. Figure \ref{CatON}(a) shows the calculation for cat states~(\ref{CatState}). In this case, we reach fidelities over $90\%$ in about $60$ measurements. Moreover, Fig.~\ref{CatON}(b) shows results similar to the qubit case given by Fig. (\ref{Graf2}), surpassing fidelities of $90\%$ in less than $40$ iterations. The last figure reflects the fact that, for the state in Eq.~(\ref{ONState}), the protocol is reduced to the qubit case, given that only two states are involved in the superposition. Thus, all states of the form in Eq.~(\ref{ONState}) have the same performance as the qubit case. Therefore, this state can be considered analog to the qubit state for high dimensions.

We can see from Figs. \ref{Graf10}, \ref{Coherent10}, and \ref{CatON}, that the learning speed is inversely proportional to the parameter $\epsilon$, which means that a small value of $\epsilon$ implies a rapid increase in fidelity between $A$ and $E$; that is, it increases the speed of learning. On the other hand, the maximum learning is also directly proportional to $\epsilon$; in other words, a small value of $\epsilon$ means lower maximum fidelities between $A$ and $E$. Also, these examples show that our proposal can be extended to high-dimensional systems. It is pertinent to emphasize that our protocol for qubit and multilevel cases employs two-level operators $\mathcal{U}_A^{(k)}$, and each iteration only needs to calculate the operator $\mathbb{U}^{(k)}=\mathcal{U}^{(k-1)}\mathbb{U}^{(k-1)}$. Hence, the protocol does not need to store the complete agent history, which is an advantage in terms of the required resources.

This protocol can be implemented in any platform that enables the logical operator $U^{NOT}_{a,b}$ for qubits, or $U^{XOR}_{a,b}$ for qudits, and digital feedback loops, as is the case of circuit quantum electrodynamics (cQEDs). This platform has particular relevance due to its fast development in quantum computation~\cite{Blais2004,Devoret2004,Hofheinz2008,Hofheinz2009,DiCarlo2009,Devoret2013,Otterbach2017}. Current technology in cQEDs allows for digital quantum feedback loops with elapsed times of about $2 \mu s$ and fidelities around $99\%$~\cite{Riste2012A,Riste2012D}, well-controlled one and two-qubits gates with fidelities over $99\%$ in less than $1[\mu s]$~\cite{Barends2016}, with qubits with coherence times about $100[\mu s]$~\cite{Paik2011,Rigetti2012}. This allows for more than $20$ iterations of our protocol, a sufficient number for a feasible implementation. Additionally, in the past decade, multilevel gates have been theoretically proposed~\cite{Strauch2011,Mischuck2013,Kiktenko2015}, and efficient multiqubit gates have recently been proposed using a ML approach \cite{Zahedinejad2015,Zahedinejad2016}, providing all the necessary elements for the experimental implementation of the general framework of this learning protocol. 

\section{Conclusions}
We propose and analyze a quantum reinforcement learning protocol to adapt a quantum state (the agent) to another, unknown, quantum state (the environment), in the context where several identical copies of the unknown state are available. The main goal of our proposal is for the agent to acquire information about the environment in a semiautonomous way, namely, in the reinforcement learning spirit. We show that the fidelity increases rapidly with the number of iterations, reaching for qubit states average fidelities over $90\%$ with less than $30$ measurements. Also, for states with dimension $d>2$, we obtain an average fidelity of over $80\%$ for $d=11$, with about $400$ measurements, which shows that our proposal works in the case of a large Hilbert space giving a scalability potential. The performance is improved for special cases such as coherent states (average fidelities of $85\%$ with less than $100$ iterations), cat states (average fidelities of $90\%$ with about $60$ iterations), and states of the form $(\ket{0}+\ket{n})/\sqrt{2}$ (average fidelities of $90\%$ with less than $40$ iterations).

The performance of the protocol is handled by the value of the parameter $\epsilon$ and by the number of states involved in the superposition of the environment state $ E $ in the measurement basis. For a small $\epsilon$ we get a high learning speed and a reduced maximum learning. Moreover, the number of states in the superposition is related to the overall performance of the protocol; that is, a superposition of fewer terms provides better performance, which increases learning speed as well as maximum learning, requiring less iterations to obtain high fidelity. These two facts imply that a possible improvement of the protocol can be achieved by using a dynamic parameter $\epsilon$ and a measurement device that can change its measurement basis throughout the protocol to reduce the number of states involved in the overlap of the state $E$. Besides, since our protocol increases the fidelity with a small number of iterations, it is useful even when the number of copies of $E$ is limited. Finally, this protocol opens up the door to the implementation of semiautonomous quantum reinforcement learning, a next step for achieving quantum artificial life.

The authors acknowledge support from CONICYT Doctorado Nacional 21140432, Direcci\'on de Postgrado USACH, Financiamiento Basal para Centros Cient\'ificos y
Tecnol\'ogicos de Excelencia (Grant No. FB0807), Ram\'on y Cajal Grant RYC-2012-11391, MINECO/FEDER FIS2015-69983-P and Basque Government IT986-16.


%
	
\end{document}